\documentclass{tPHM2e}

\usepackage{epsfig}

\begin{document}

\doi{10.1080/1478643YYxxxxxxxx}
\issn{1478-6443}
\issnp{1478-6435}
\jvol{00} \jnum{00} \jyear{2009} \jmonth{1 January}

\markboth{R. Spatschek, E. Brener, and A. Karma}{Philosophical Magazine}

\articletype{Review Article}

\title{Phase field modeling of crack propagation}

\author{Robert Spatschek$^{\rm a}$$^{\ast}$\thanks{$^\ast$Corresponding author. 
Email: robert.spatschek@rub.de \vspace{6pt}}, Efim Brener$^{\rm b}$, and Alain Karma$^{\rm c}$\\
\vspace{6pt} $^{\rm a}${\em{Interdisciplinary Center for Advanced Materials Simulation (ICAMS), Ruhr University 44780 Bochum, Germany}}; \\
$^{\rm b}${\em{Institute for Solid State Physics, Research Center 52425 J\"ulich, Germany}}; \\
$^{\rm c}${\em{Physics Department and Center for Interdisciplinary Research on Complex Systems, Northeastern University, Boston, MA 02115, USA}}\\
\vspace{6pt}\received{\today} }




\maketitle

\begin{abstract}
Fracture is a fundamental mechanism of materials failure. Propagating cracks can exhibit a rich dynamical behavior controlled by a subtle interplay between microscopic failure processes in the crack tip region and macroscopic elasticity. We review recent approaches to understand crack dynamics using the phase field method. This method, developed originally for phase transformations, has the well-known advantage of avoiding explicit front tracking by making material interfaces spatially diffuse. In a fracture context, this method is able to capture both the short-scale physics of failure and macroscopic linear elasticity within a self-consistent set of equations that can be simulated on experimentally relevant length and time scales. We discuss the relevance of different models, which stem from continuum field descriptions of brittle materials and crystals, to address questions concerning crack path selection and branching instabilities, as well as models that are based on mesoscale concepts for crack tip scale selection.
Open questions which may be addressed using phase field models of fracture are summarized.
\end{abstract}

\date{\today}

\section{Introduction}

The dynamics of crack propagation is an important and long standing challenge in materials science and solid-state physics \cite{Lawn:1993fp, marder:24}, and in the recent years the physics community saw a rebirth of interest in the problem of dynamic fracture, also in combination with the concept of phase field modeling \cite{Langer:1986fk, PhysRevB.31.6119, Chen:2002qr, Karma:2005uq, Steinbach:2009nx}.
In this respect, fracture is yet another extension of a concept that was originally introduced to describe and simulate the kinetics of phase transitions -- as the historical name ``phase field'' suggests.
Today, this methodology has entered an ``inflationary stage'', and it has been successfully extended to various new applications in physics, materials science, but also biology and medicine.
The characteristic property of phase field methods is the existence of one or more ``phases'', for example solid or liquid phases, which can be transformed into each other. While a scalar phase field was originally introduced to distinguish between different phases \cite{Langer:1986fk,PhysRevB.31.6119}, other fields have been introduced
to characterize other properties such as the local crystal orientation \cite{WKLC:2003}.
On this broader scope, the phase field variables are therefore more appropriately denoted as order parameters.
For the modeling of cracks, such an order parameter distinguishes between the solid phase and the ``broken'' state inside the crack.
As usual in the phase field context, the order parameter changes smoothly between the states at the crack surfaces.
The growth of a crack becomes then conceptually comparable to the front propagation in a first order phase transition.
During the past years, several important questions concerning the growth of cracks have been successfully tackled by this new paradigm.

The uniform motion of a crack is relatively well understood in the framework of continuum theories \cite{Freund:1998yq, Broberg:1999vn, Rice:1968kx}.
Here, the conventional approach is to treat the  crack as a front or interface separating broken and unbroken regions of the material;
propagation is governed by the balance of the elastic forces in the materials and cohesive stresses near the crack tip \cite{PhysRevLett.76.1497, PhysRevLett.82.2314, PhysRevLett.79.877}.
Many characteristic features of crack propagation are nowadays well established by experimental studies \cite{PhysRevLett.82.3823, PhysRevLett.74.5096, PhysRevLett.76.2117, PhysRevLett.67.457, PhysRevE.53.5637, Sharon:1999ly, 0295-5075-30-6-004, BoudetCilibertoSteinberg1996}.
As soon as the flux of energy to the crack tip exceeds a critical value, the crack becomes unstable and starts to branch while emitting sound waves.
These phenomena are consistent with the continuum theory of sharp crack tips, but it fails to describe it, because the details of crack growth, in particular the chosen crack path and velocity, depend on details of cohesion on microscopic scales \cite{Fineberg:1999zr}.
Nevertheless, empirical energy balances and simple propagation laws that are frequently used in engineering applications, cannot account for the richness of actual fracture phenomena. In particular, they cannot predict
dynamical instabilities of fast moving cracks. The fundamental mechanisms of those instabilities have been extremely difficult to elucidate because they appear to result from a non-trivial coupling between dynamical phenomena inside the crack tip region, known as {\em process zone}, and linear elasticity, with no clear separation of scale between atoms and the system size. 
 
Large scale molecular dynamics (MD) simulations with about $10^7$ atoms allowed to get deeper insights into the growth behavior of cracks \cite{PhysRevLett.77.869, PhysRevLett.78.479, PhysRevLett.78.689, PhysRevLett.79.1309, DynFracSpecIssue1999}.
Although limited to submicron samples and very short timescales, these simulations were able to reproduce key features of crack propagation like the initial acceleration and the onset of instabilities.
Nevertheless, a detailed understanding of the complex physics of crack propagation, in particular aspects of the pattern formation process, still remain a major challenge \cite{AlexanderHellemans08141998}.

At this level, continuum descriptions, in particular phase field methods that avoid dynamical artifacts which are associated with the breaking of translational and rotational symmetry \cite{PhysRevE.68.036118, Marder19951}, offer a useful and complementary perspective on crack propagation as a pattern formation process.
The focus of this article is on phase field formulations. Fracture has also been modeled by level set approaches, see e.g.~\cite{Chopp:2002uq}.
Furthermore, cracks are only a particular type of defects in materials;
for a recent overview on the phase field modeling of defects in general see \cite{WangDefect2010}.

\section{Basic aspects of fracture}

\begin{figure}
\begin{center}
\includegraphics[trim=0cm 0cm 10cm 0cm, width=6cm]{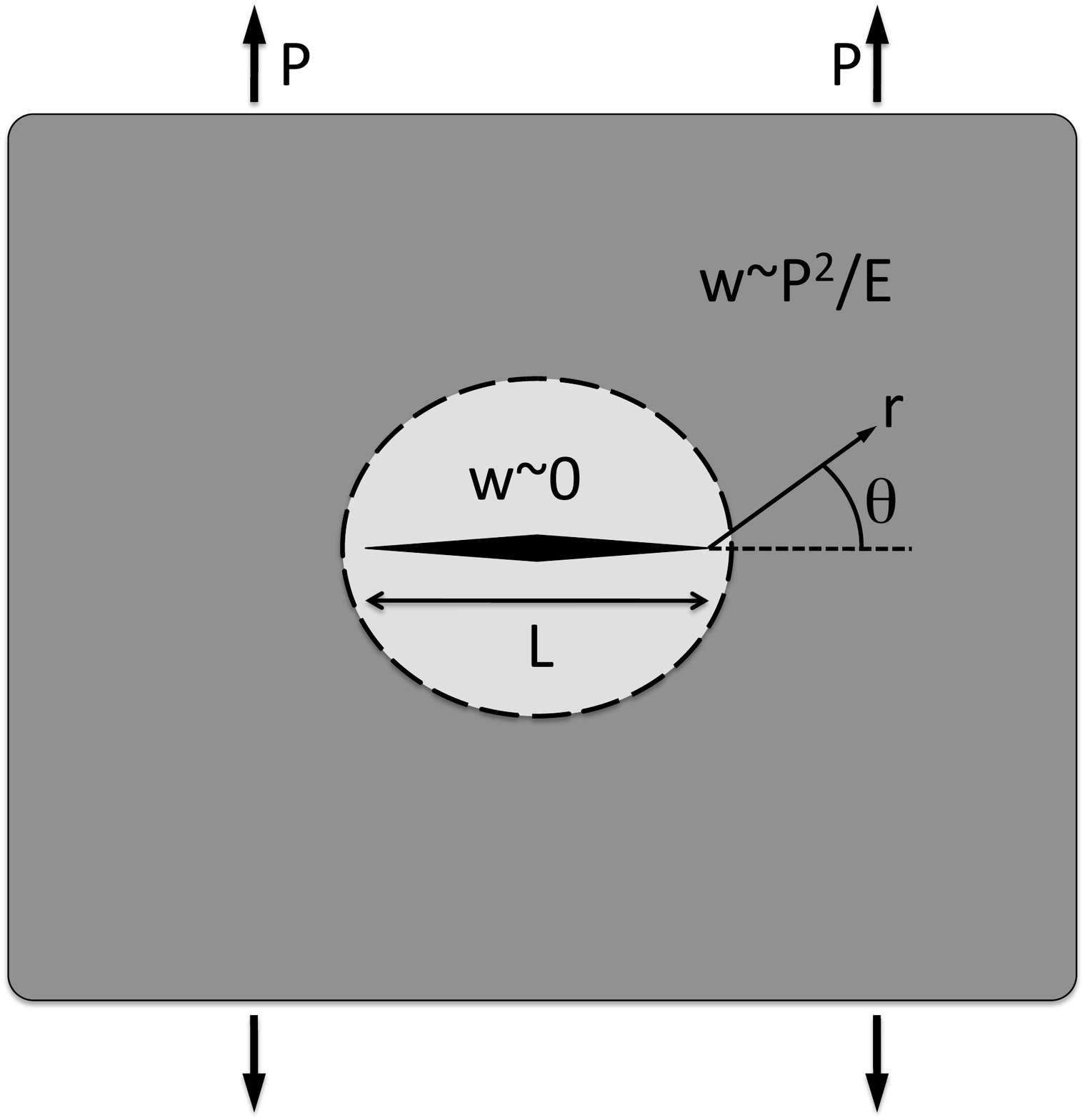}
\caption{Stress relaxation and interface creation due to the appearance of a crack. Near the (sharp) crack tip the stress becomes singular in the vicinity of the crack in the framework of the linear theory of elasticity. \label{singularity}}
\end{center}
\end{figure}

One of the cornerstones of fracture in brittle materials is the Griffith criterion \cite{Griffith:1920hl}.
Materials under tension store elastic energy with density $w\sim P^2/E$ with the applied stress $P$ and an elastic modulus $E$.
The appearance of a crack with length $L$ leads to a local relaxation of the elastic strain, $\Delta F_{el}\sim wL^2$ (here in two dimensions), but at the same time the new crack surfaces are created, which lead to an additional surface, or more generally, fracture energy, $\Delta F_s \sim \gamma L$, see Fig.~\ref{singularity}.
For short cracks, the increase of the interfacial term is most important, whereas for long cracks the second term dominates;
hence fracture is similar to nucleation in first order phase transitions, eventually with a cohesive interaction between the crack surfaces \cite{PhysRevLett.63.171}.
The critical length of a crack, $L_G\sim E\gamma/P^2$, is called the {\em Griffith length}.

Near the crack tip the stress has a universal singular behavior in the framework of the linear theory of elasticity \cite{William:1957qd, Irwin:1957qr}, see Fig.~\ref{singularity},
\begin{figure}
\begin{center}
\includegraphics[width=10cm]{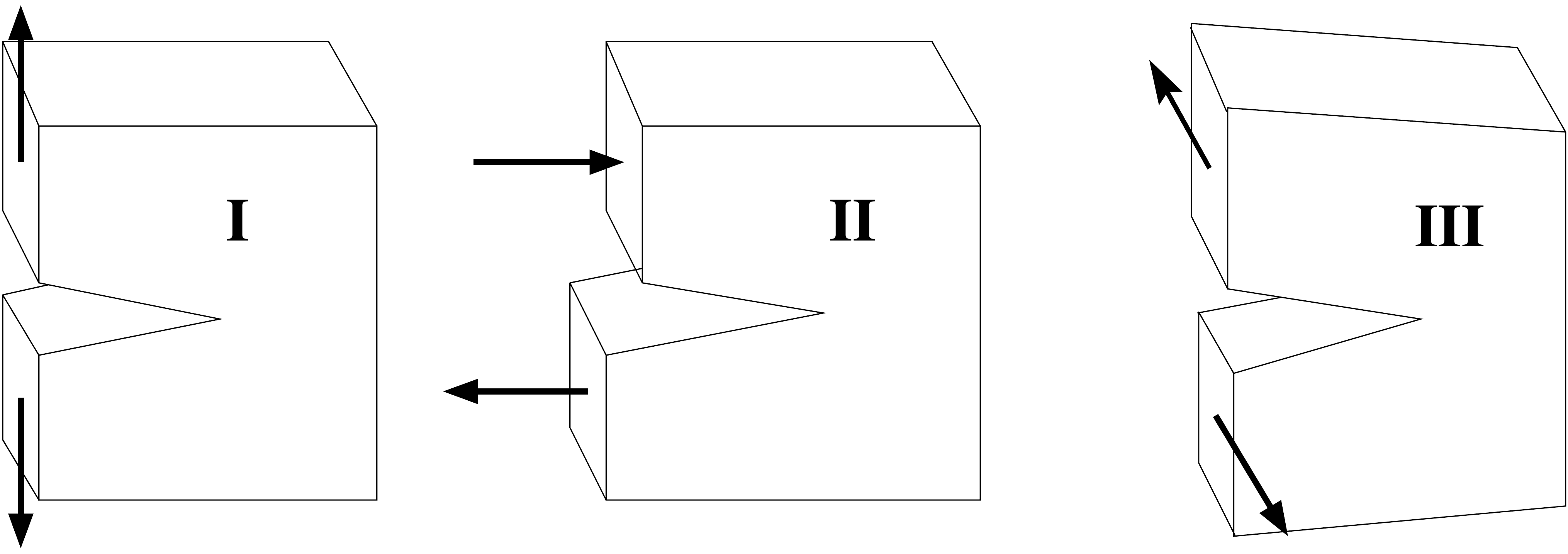}
\caption{The different loading modes of a crack. \label{modes}}
\end{center}
\end{figure}
\begin{equation} \label{eq1}
\sigma_{ij}^m = \frac{K_m}{\sqrt{2\pi r}} f_{ij}^m(\theta),
\end{equation}
where $K_m$ are the stress intensity factors for the three modes of loading (see Fig.~\ref{modes}), and $\theta$ is the angle between the radial vector of length $r$ with origin at the (sharp) crack tip and the local crack direction.
The angular depedences $f_{ij}^m(\theta)$ are known functions, see e.g.~\cite{Freund:1998yq, Broberg:1999vn, Rice:1968kx}.
We note that the strength of this singularity, $\sigma\sim r^{-1/2}$, differs from dislocations, where stresses scale as $\sigma\sim r^{-1}$.
In both cases, the cutoff for the singularity and the regularization of stresses is important since stresses cannot be infinite \cite{PhysRevB.70.104104};
this can be due to nonlinear phenomena in the process zone or a finite crack tip radius.

For propagation of a crack, the energy release rate, also known as crack extension force,
\begin{equation}
G = \frac{1-\nu^2}{E} (K_I^2 + K_{II}^2) +\frac{1}{2\mu} K_{III}^2
\end{equation}
has to exceed a material dependent threshold $G_c$;
this is equivalent to the Griffith criterion.
In the ideal case, this energy barrier is exactly twice the surface energy of the material, $G_c=2\gamma$, since two new surfaces are created;
in reality however, this threshold is usually much higher.

\section{Continuum field models}

The main aim of continuum field models of fracture is to provide a unified framework for the whole phenomenology of fracture, ranging from crack initiation to oscillations and branching.

Probably the first model in a series of phase field like descriptions of crack growth in brittle materials was developed by Aranson, Kalatsky and Vinokur \cite{PhysRevLett.85.118}, focusing on mode I cracks in two dimensions.
Apart from the displacement field, which obeys the elastodynamic equations together with a damping term, they introduce an order parameter $\rho$ which accounts for the dynamics of defects.
This order parameter has the value $\rho=1$ in the solid, whereas it vanishes inside the cracks, $\rho=0$.
It is assumed that this order parameter changes smoothly at the crack surfaces on a scale that is large in comparison to the atomic spacing, to justify a continuum description.
This phenomenological model already captures many important features, like crack initiation, propagation, dynamic fracture instability, sound emission, crack branching and fragmentation.

In this model, the elastic modulus is assumed to be proportional to the order parameter, $E=E_0 \rho$, which implies that the interior of an ideal crack is stress free.
The stress is given by
\begin{equation}
\sigma_{ij} = \frac{E}{1+\nu} \left( \epsilon_{ij} + \frac{\nu}{1-\nu} \epsilon_{ll}\delta_{ij} \right) + \chi \dot{\rho} \delta_{ij},
\end{equation}
where the last term mimics a hydrostatic pressure due to the creation of new defects.
The elastodynamic equation contains a viscous damping
\begin{equation}
\rho_0 \ddot{u}_i = \eta \nabla^2 \dot{u}_i + \frac{\partial\sigma_{ij}}{\partial x_j}.
\end{equation}
The evolution of the order parameter is assumed to follow a relaxation law
\begin{equation}
\dot{\rho} = D\nabla^2\rho - \alpha\rho(1-\rho) [1-(b-\mu \epsilon_{ll})\rho] + c\rho(1-\rho)\frac{\partial \rho}{\partial x_l} \dot{u}_l,
\end{equation}
which contains several material parameters.
The first, diffusive term is analogous to the contribution of the gradient term in a phase field model, and similarly the second term corresponds to a double well potential with minima at $\rho=0$ and $\rho=1$.
The last term reflects an advective contribution, which couples the local order parameter to the displacement rate and is responsible for the localized shrinkage of the crack due to material motion.
It turns out that this phenomenological term is important to maintain a sharp crack tip.

For low driving forces, the model predicts steady state growth of a single straight crack in a strip, and the distribution of the elastic fields is in agreement with linear elastic fracture mechanics.
Above a critical crack velocity, which is some fraction of the Rayleigh speed $v_R$ (sound speed of a wave traveling along a free surface), branching occurs;
the threshold velocity depends on the material parameters and is roughly in the range $(0.3-0.55) v_R$.
The instability manifests itself as pronounced velocity oscillations, crack branching and sound emission from the crack tip.
A velocity gap, i.e.~a minimum crack velocity as in lattice models \cite{Fineberg:1999zr}, is not found.

A frequent and important analysis for phase field-like models for crack growth in a channel geometry is related to the asymptotic behavior in the tail region far behind the crack tip.
There (depending of course on the boundary conditions, and we assume that the strip is loaded by a fixed displacement), both the order parameter and the elastic fields become homogeneous, and therefore a theoretical analysis becomes feasible.
In the above model \cite{PhysRevLett.85.118} the crack opening depends logarithmically on the strip width $L$ for fixed applied strain.
This is in contrast to the expectation that the material should be fully relaxed in the wake, thus all displacement should be accumulated within in crack, and hence the crack opening should increase linearly with the strip width.

To overcome this problem, Karma, Kessler and Levine (KKL) proposed a different phase field approach where the scalar order parameter describes the state of the material in Lagrangian ``material coordinates'', instead of the density in Eulerian coordinates \cite{PhysRevLett.87.045501}.
They investigated in particular mode III cracks, which makes the description simpler since only the out-of-plane displacement component $u_z$ exists in a two-dimensional description.
They link the description of cracks closer to conventional phase field modeling, as it is widely used e.g.~for solidification.
The order parameter distinguishes between broken, $\phi=0$, and unbroken states, $\phi=1$, and its dynamical evolution is derived variationally from a free energy functional,
\begin{equation}
F = \int d{\mathbf x} \left[ \frac{1}{2} D_\phi (\nabla\phi)^2 + V_{DW}(\phi) + \frac{\mu}{2} g(\phi) ({\mathbf \epsilon}^2 - {\mathbf \epsilon}_c^2)\right]
\end{equation}
and $\tau\partial_t \phi=-\delta F/\delta \phi$.
The ``bulk'' states correspond to the minima of a double well potential, $V_{DW}=\phi^2(1-\phi)^2/4$, which together with the gradient term is responsible for the fracture energy.
The elastic term involves the shear modulus $\mu$ and the strain ${\mathbf \epsilon}$, which induces breaking of the material if it exceeds a critical strain ${\mathbf \epsilon}_c$.
The choice of the coupling function $g(\phi)$ turns out to be critical and should scale as $g(\phi)\sim \phi^{2+\alpha}$ with $\alpha>0$ for small values of $\phi$, in order to obtain full stress relaxation in a completely broken solid in the limit of large system sizes.
We mention the similarity of the model to conventional phase field models of solidification, where the last term corresponds to the deviation from the melting temperature $g(\phi)L(T-T_M)/T_M$;
this also shows that the crack motion is here a fully reversible process, i.e.~the crack can heal if the load is released.
The choice of the coupling function, however, is here more strict, which is due to the fact that the crack width is directly related to the phase field interface thickness, hence the model does not operate in or close to the sharp interface limit.

The model was applied to the dynamics of mode III cracks in \cite{PhysRevLett.92.245510}.
\begin{figure}
\begin{center}
\epsfig{file=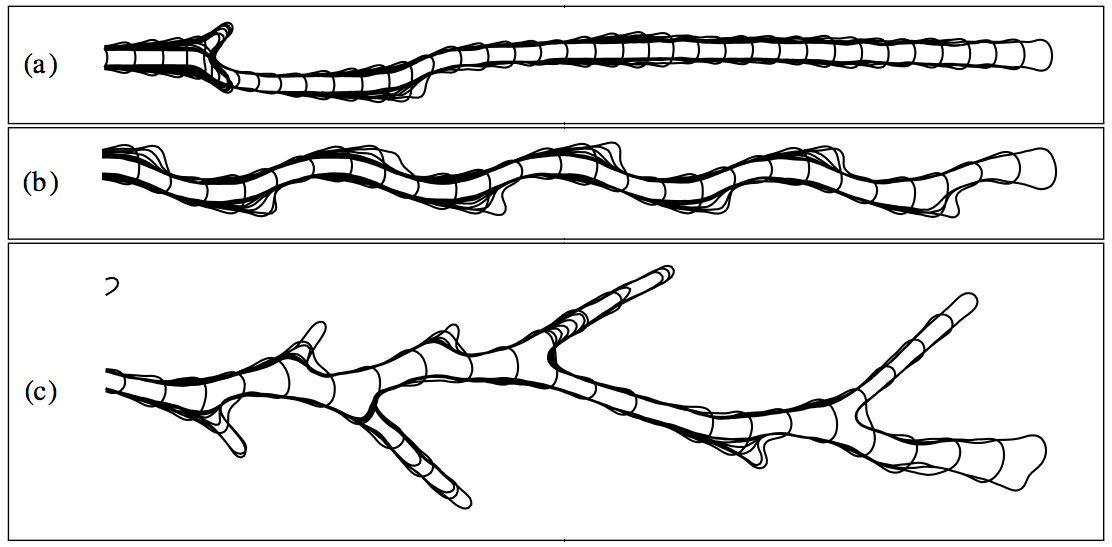, width=10cm}
\caption{Steady and unsteady propagation of mode III cracks with increasing load from (a) to (c). Straight crack propagation is stable at low velocity  (a), becomes oscillatory at intermediate velocity (b), and exhibits branching (c) at high velocity. Taken from \cite{PhysRevLett.92.245510}.}
\label{karma}
\end{center}
\end{figure}
If the relaxation timescale $\tau$ is long, the growth of the crack is slow and controlled by the interface kinetics;
in this limit the elastic fields are quasistatic.
In contrast, for fast growth the crack speed is limited by the sound speed, and all dissipation in this brittle limit takes place in the interface region.
Three different basic regimes of solutions are found, see Fig.~\ref{karma}.
Steady state growth for low driving forces, an asymmetric tip-splitting mode for intermediate loads, where the crack tip follows a snake-like sinusoidal trajectory, and finally a chaotic tip-splitting regime with well-developed branches for large loads.
The maximum velocity of a single straight crack in the inertial limit turns out to be $v_{max} \approx 0.41 c_s$ ($c_s$ is the shear wave speed), and it is argued that the limitation of the crack speed is due to tip blunting as a result of relativistic contraction;
the observed branching angles close to the onset of branching are similar to analytical predictions \cite{AddaBedia2004}.

Henry and Levine extended the model further to two-dimensional plane strain situations and investigated the growth of cracks under mode I and mode II conditions \cite{PhysRevLett.93.105504}, recovering the same basic dynamical behavior of the previous mode III phase field study of fast moving cracks. 
Also, they generalize the elastic energy and also suggest a symmetry breaking term to go beyond situations where the elastic energy is quadratic in strain, where the situation is symmetric under tension and compression without advective contributions, implying that cracks also grow in compressed materials.
Therefore, Henry and Levine generalize the contribution of the elastic energy in the free energy functional, that is used to express the dissipative phase field dynamics, $g(\phi)(E_\phi - E_0)$ ($E_0$ corresponds to $\epsilon_c^2$ in the mode III model), to
\begin{equation}
E_\phi = \left\{ 
\begin{array}{cc}
\frac{1}{2}\lambda\epsilon_{ii}^2 + \mu \epsilon_{ij}^2 & \mathrm{if}\, \mathrm{tr}\, \epsilon > 0, \\
\frac{1}{2}\lambda\epsilon_{ii}^2 + \mu \epsilon_{ij}^2 -a K\epsilon_{ii}^2 & \mathrm{if}\, \mathrm{tr}\, \epsilon > 0
\end{array}
\right.
\end{equation}
with the modulus of compression $K=(\lambda+\mu)/2$ and $a>1$ is a parameter that prevents breaking under compression.
This avoids situations where additional shear forces on top of a mode I loading lead to splitting of the crack and symmetrical growth of the two branches around the previous propagation direction.
Instead, the material breaks now only in the tensile region.
For higher loading also the mode I crack becomes unstable, and secondary cracks develop out of a branching instability.
Notice, that not only the symmetry breaking model by Henry and Levine but also the original KKL model satisfy the principle of local symmetry, i.e. a reorientation of the crack growth direction such that the mode II stress intensity factor vanishes, $K_{II}=0$.
This property will be discussed in more detail in section \ref{path}.

The model was then applied to biaxial loading (an additional tensile stress component acts in the growth direction), where a supercritical Hopf bifurcation from straight to sinusoidal oscillatory is observed.
A similar phenomenon was previously observed experimentally in rubber \cite{PhysRevLett.88.014304}.
Remarkably, this instability, which may be interpreted as the attempt of the system to get rid of the additional elastic energy related to the longitudinal strain, is observed only for dynamical crack growth but not for overdamped elasticity;
the wavelength of the instability depends linearly on the strip width that is used for the simulation.

A recent and more detailed investigation of the model, especially concerning the onset of branching, has been done in \cite{0295-5075-83-1-16004}, and a reasonable agreement of critical velocities and branching angles in experiments \cite{PhysRevLett.74.5096, PhysRevLett.76.2117, PhysRevB.54.7128} and theoretical predictions \cite{AddaBedia2005227, 2494202220070201} has been found.
Also, the dynamics are altered in comparison to the preceding models such that $\partial_t\phi<0$, which means that cracks can only grow and not recede. 
Therefore, sidebranches do not disappear even far behind the  crack tip, otherwise they would shrink in order to minimize the interfacial energy.
The same concept, together with the use of nonlinear elastic laws, is briefly presented in \cite{Kuhn:2008rc}.

Marconi and Jagla presented a diffuse interface approach to slow fracture in brittle materials using a different concept \cite{PhysRevE.71.036110}.
They do not introduce an additional order parameter to discriminate between broken and unbroken regions, but use the strain field itself as ``order parameter''.
The description of material failure is here entirely through the nonlinear form of the elastic free energy density, which saturates for high strains, therefore representing the broken material inside the crack;
the limiting value of the nonlinear stress-strain relation determines the fracture energy.
A similarity to phase field models stems from the presence of a gradient energy term, which acts here on the components of the strain tensor, thus introducing also second order derivatives of the strain in the equations of motion (in other models, the strain appears in the phase field evolution equation only locally and with first derivative in the elastic equation).
Within this model, the fracture energy depends on the load, and ad hoc modifications of the model are necessary for corrections, which introduce additional parameters to the theory.

Most phase field models for fracture use nonconserved order parameters, and exceptions are rare.
One of them is the work by Eastgate et al. \cite{PhysRevE.65.036117}.
In their case, the phase field is interpreted as the (normalized) mass density of the material surrounding the crack, similar to \cite{PhysRevLett.85.118}.
Inside the crack, it becomes zero, whereas in the solid it is $\phi=1-\epsilon_{kk}$, which takes into account the local compression or expansion of the material due to elastic strain.
Both the dynamics of the phase field and the elastic displacement are derived from a free energy functional, and the order parameter evolves according to $\dot{\phi}=-\nabla\cdot{\mathbf j}$, where the flux $\mathbf{j}$ is not only driven by the energy decay, but also contains the advective contribution due to the elastic displacement.
The displacement field obeys a viscous law, and is therefore mostly appropriate for the description of fracture e.g.~in colloidal crystals.
This is one of the main differences in comparison to the work of Bhate et al.~\cite{bhate:1712} for modeling of stress voiding in electromigration, where quasistatic elasticity is assumed.
The model \cite{PhysRevE.65.036117} is used to describe the deformation of an initially spherical hole in a two-dimensional strained system and the evolution towards a crack.
Limitations of the model are related to the fact that inside the crack the phase field does not fully reach the vacuum state $\phi=0$, but retains a value that is proportional to the tangential strain along the crack surfaces.
Related to this artifact is a slight deviation from the correct value of the Griffith point.

\section{Crack path prediction}
\label{path}

The prediction of the path that a crack chooses while it propagates through a brittle material has been a long standing problem in fracture mechanics.
In a conventional picture, the equations of linear elasticity are solved for traction free crack surfaces with sharp crack tips \cite{Broberg:1999vn}.
Although the Griffith condition provides a criterion for crack growth, it is insufficient to predict the curvilinear crack paths or crack kinking, or even branching angles.
The generally accepted condition $K_{II}=0$ assumes that the crack propagates in such a direction that it is in a pure opening-mode state with a symmetrical stress distribution about its local axis \cite{Goldstein:1974yq}.
This {\em principle of local symmetry} has been rationalized using plausible arguments in the classical framework of fracture mechanics \cite{Cotterell:1980vn,Oleaga20012273} and has also been argued from the continuity of the chemical potentials at the fracture tip \cite{PhysRevLett.81.5141}.
Variational formulations based on energy minimization were considered in \cite{Francfort19981319, Bourdin2000797} as a proposal to overcome the limitations of Griffith's theory.
Hodgdon and Sethna suggested equations of motion for a crack based on symmetry arguments \cite{PhysRevB.47.4831}.
Nevertheless, crack propagation laws are hard to derive without a detailed knowledge of the physics of the process zone, where the elastic energy is transformed into new crack surfaces and dissipated.
In addition, symmetry considerations cannot be invoked when the fracture energy and/or elastic properties become anisotropic \cite{0295-5075-66-3-364}.

Hakim and Karma used the KKL phase field model \cite{PhysRevLett.87.045501} to derive the principle of local symmetry in a fairly generic way, and provide even generalized conditions to predict crack paths \cite{PhysRevLett.95.235501, Hakim2009342}.
Similar to considerations in classical fracture mechanics \cite{Rice:1968kx}, they calculate the energy flux into the crack tip.
Apart from a dissipative term, it consists of purely elastic contributions, which are conservative and can therefore be expressed by the energy flux through a circle around the crack tip;
since its radius can be large, this contribution to the energy -- excluding the contribution from the segment that crosses the crack surfaces -- is described by the singular stress field and yields the same result as Rice's $J$ integral \cite{Rice:1968rt}, which is the crack extension force $G$, and also Eshelby's torque $G_\theta=dG(\theta)/d\theta$, where $G(\theta)$ is the extension force for a slightly kinked crack into direction $\theta$ \cite{Eshelby:1975ys}.
This torque term tends to turn the crack in a direction that maximizes the crack extension force.
The crucial difference to classical fracture mechanics is that the phase field model also contains a description of the process zone, and the integral contribution along a line that crosses the crack gives contributions from cohesive forces, thus being related to the crack surface energy.
A similar torque term gives a contribution proportional to the Herring torque $\gamma_\theta=d\gamma/d\theta$ \cite{Herring:1951fr}, and therefore turns the crack into a direction that minimizes the surface energy in anisotropic media.
The main result for quasistatic fracture is the generalized principle of local equilibrium, stating
\begin{equation}
K_{II} = -\frac{[2\gamma_\theta(0) + f_2]E}{2(1-\nu^2)K_1},\label{K2cond}
\end{equation}
where $f_2$ is the dissipative force perpendicular to the crack, which vanishes identically in isotropic media and otherwise becomes negligible close to the Griffith point.
For isotropic media, $\gamma_\theta=0$, the prediction recovers therefore the anticipated relation $K_{II}=0$, otherwise a shear loading balances the Herring torque term.

\begin{figure}
\begin{center}
\includegraphics[width=8cm]{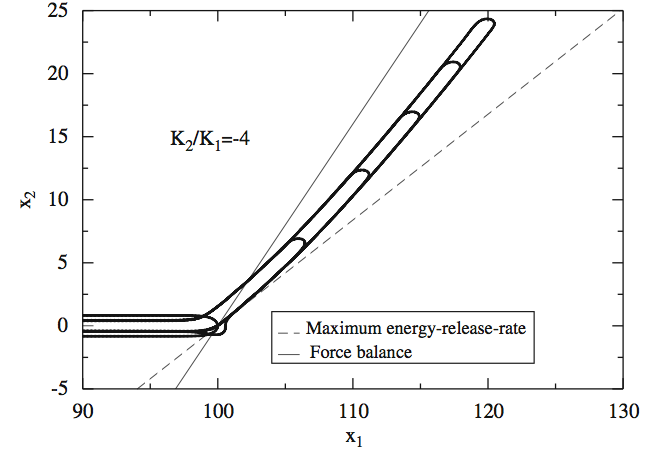}
\caption{Comparison of phase field simulations and theoretical predictions of the maximum energy release rate criterion and the force balance condition derived from the phase field model for kinked cracks in a material with an anisotropic surface energy. Taken from \cite{Hakim2009342}.}
\label{kink}
\end{center}
\end{figure}

The principle of local symmetry and an alternative suggestion, the principle of maximum energy release rate, give identical results for smooth curvilinear cracks and only small differences for kinks \cite{Cotterell:1980vn, Hutchinson:1992rc, Amestoy:1992rw}, and thus it is difficult the discriminate between the predictions.
In anisotropic media, however, these principles can give substantially different predictions for the kink angles for specific surface energy anisotropy and loading conditions.
The phase field simulations in \cite{Hakim2009342} show that the kink cracks emerge from the main crack tip at an angle, which is initially close to the prediction of the maximum energy release rate criterion, but approaches the angle predicted by the force balance condition (generalized principle of local symmetry) on the scale of the process zone, see Fig.~\ref{kink}.

For antiplane shear, where only one stress intensity factor exists, the prediction of the crack path requires consideration not only of the singular stress term, but also of the following, subdominant contribution in the stress field expansion, and the principle of local symmetry demands here that its prefactor becomes zero \cite{Barenblatt:1961db}.

The problem of crack path prediction is also essential for the case of thermal loading, and has been investigated intensively theoretically and experimentally.
Typically, a glass strip with a notch at its end is pulled at a constant velocity from an oven into a cold bath.
The control parameters in this experiment are mainly the width of the strip, the temperature gradient between the oven and the cold bath, and the pulling velocity.
If the velocity is small enough, a crack does not propagate.
Above a first critical velocity, the crack starts propagating following a straight centered path, and above a second critical velocity, the crack begins to oscillate with a well defined wavelength.
For a recent investigation, including a phase field model for this problem, which shows that the transition from straight to oscillatory cracks is supercritical, we refer to \cite{corson-2009-158} and references therein.

\section{Phase field modeling of fracture in crystalline materials}

Crack propagation in crystalline materials is inherently anisotropic because the fracture energy depends on the orientation of the fracture surface with respect to some underlying set of crystal axes. For this reason, cracks are often observed to propagate along low energy cleavage planes.
Crystalline anisotropy can in principle be included in the KKL phase field model \cite{PhysRevLett.87.045501} by modifying square gradient terms in the free-energy density to break rotational invariance \cite{PhysRevLett.95.235501,Hakim2009342}.  Using the general propagation law for crack paths in anisotropic materials given by Eq. (\ref{K2cond}), Hakim and Karma have derived a condition for a crack to escape from a cleavage plane. This condition stems from the physical picture that a crack will be ``trapped'' along a low energy plane corresponding to a cusp in the $\gamma$-plot (i.e. the plot of surface energy as a function of orientation) until the Eshelby configurational torque and hence $K_{II}$ is large enough to rotate the crack away from this plane. 

Another independent development of crack propagation in a phase field context has been developed by Jin, Wang and Khachaturyan \cite{Jin:2003dn} in the spirit of the phase field microelasticity approach, that had previously been used to study e.g.~coherent precipitation and ordering \cite{Wang:1993zl, LeBouar19982777, Hu2001463}, martensitic transformations \cite{Wang1997759, Jin20012309} and dislocation dynamics \cite{Jin:2001rm, Wang20011847}.
The main step is here the solution of the elastostatic problem, since the discontinuous system, that contains cracks and voids, is not elastically homogeneous.
Instead the elastic modulus of the material becomes equal to zero within the inclusions.
For the solution of this problem beyond a perturbative treatment \cite{PhysRevB.52.15909}, the material is equivalently considered  as a continuous homogeneous body with a mesoscopically heterogeneous misfit-generating stress-free strain $\epsilon_{ij}^0$, also know as eigenstrain.
If the misfit strain is known everywhere, the elastic problem is solved by minimizing the free energy $E_{el}$ with respect to the displacements.
Here, the eigenstrain is an additional degree of freedom that is present only inside the cracks or voids.
Additional minimization with respect to the eigenstrain leads to a stress-free state inside the inclusions \cite{wang:1351}.
This is conceptually similar to the representation of a crack as pile-up of dislocations \cite{Bilby:1968kx}, and the eigenstrain $\epsilon_{ij}^0(\mathbf{r})$ plays the role of a long-lange order parameter.
It can be determined by a time-dependent Ginzburg-Landau equation, which is first applied inside the cracks and voids only,
\begin{equation}
\frac{\partial \epsilon_{ij}^0(\mathbf{r}, t)}{\partial t} = - L_{ijkl} \frac{\delta E_{el}}{\delta \epsilon_{kl}^0(\mathbf{r}, t)}
\end{equation}
with positive definite Onsager coefficients $L_{ijkl}$.

To include also a dynamical evolution of the cracks, it is assumed that the cracks can develop along cleavage planes $\alpha=1\ldots p$ with normal vectors $\mathbf{H}(\alpha)$.
The stress-free transformation strain is then assumed to have the structure 
\begin{equation}
\epsilon_{ij}^0(\mathbf{r}) = \sum_{\alpha=1}^p h_i(\alpha, \mathbf{r}) H_j(\alpha),
\end{equation}
where the $h_i$ play now the role of the phase fields.
An additional local energy term, which depends on the order parameters $h_i$ represents a cohesive energy between the crack lips and can be tuned to the desired model.
Similar to conventional phase field models, a gradient energy term is added,
\begin{equation}
E_{grad} = \sum_{\alpha=1}^p \int_V f_{grad}(\mathbf{h}(\alpha, \mathbf{r})) d\mathbf{r},
\end{equation}
and the gradient energy density reads in the simplest case that the crack tip energy does not depend on the direction of the crack front in the cleavage plane,
\begin{equation}
f_{grad} = \beta(\alpha) [\mathbf{H}(\alpha) \times \nabla (\mathbf{h}(\alpha, \mathbf{r})\cdot \mathbf{H}(\alpha))]^2
\end{equation}
with material parameters $\beta(\alpha)$.
The evolution of the long-range order parameters follows then a usual relaxation equation from the total free energy $F$
\begin{equation}
\frac{\partial h_i(\alpha, \mathbf{r}, t)}{\partial t} = - L_{ij} \frac{\delta F}{\delta h_j(\alpha, \mathbf{r}, t)} + \xi_i(\alpha, \mathbf{r}, t)
\end{equation}
with an additional Langevin noise term $\xi_i$.
This shows the conceptual similarity to models for dislocation dyamics \cite{Jin:2001rm, Wang20011847}, where the role of the functions $\mathbf{h}$ is played by the Burgers vector.
In contrast to the preceding phase field models, this description leads to crack paths that follow the cleavage planes in a crystal, and therefore do not obey the principle of local symmetry as in isotropic materials.

The interactions of dislocations and martensites with free surfaces, voids and cracks have been demonstrated to work robustly in the framework of these models e.g.~in \cite{Wang:2003qy, jin:3071, wang:6435}.
An extension of this model to phase field simulations of crack tip domain switching in ferrorelectrics is presented in \cite{0022-3727-40-4-040, Wang:2007fj}.
These materials are widely used to fabricate sensors and transducers, but tend to break for high electric or mechanical loading due to their brittleness.
Hydrate precipication and delayed hydrate cracking in zirconium are considered in \cite{HydrogenDiffusion2006};
here hydrogen or deuterium diffuse along a stress gradient towards a crack tip, where hydrides form and grow, and finally the crack develops through the hydride.

Crack propagation has also been simulated using the phase field crystal model inspired from classical density functional theory, which resolves spatially the crystalline density field \cite{PhysRevE.70.051605}.
\begin{figure}
\begin{center}
\epsfig{file=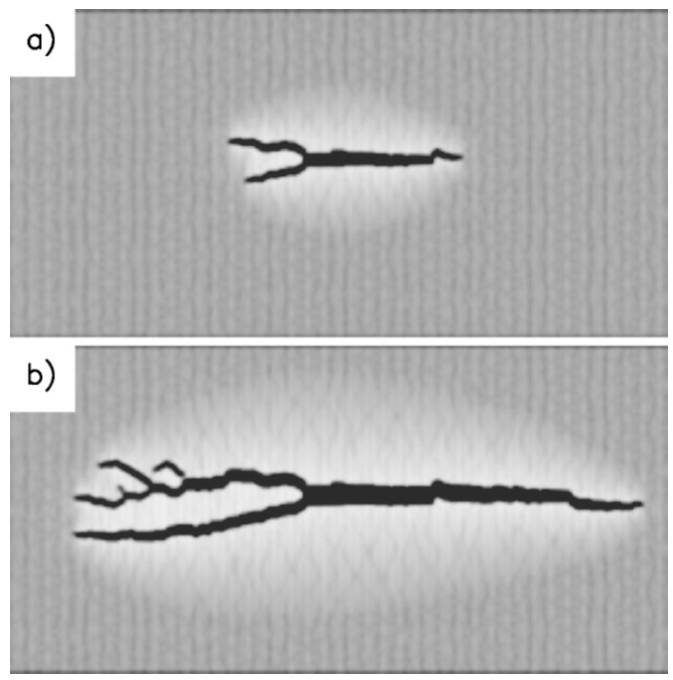, width=6cm}
\caption{Crack propagation using the phase field crystal model. The greyscale illustrates the energy density and visualizes the stress relaxation in the vicinity of the crack. Taken from \cite{PhysRevE.70.051605}.}
\label{elderfig}
\end{center}
\end{figure}
Since the phase field crystal model naturally describes dislocations, it could potentially be used to investigate the transition from ductile to brittle crack propagation. However,
further investigations will be necessary to elucidate the applicability of the model for fracture, which remains almost completely unexplored beyond pictorial examples of crack branching shown in Fig.~\ref{elderfig}.
One limitation of the original phase field crystal model is the restriction to only one (diffusive) timescale, which limits not only the interface velocity but also the elastic or plastic relaxation in the bulk.
This is appropriate for slowly propagating cracks, but is restricted in applicability for higher velocities, since then elastic waves should travel with the speed of sound instead of relaxing diffusively.
A potential method to decouple these processes is to use an acceleration term in the phase field crystal evolution equation, as demonstrated in \cite{stefanovic:225504}.

\section{Models with sharp interface limit}

Other phase field models of fracture have been developed from a different viewpoint than the models described so far in this article. Closer to the original motivation for introducing the phase field model in the context of phase transformations, this class of models uses the phase field method as a numerical tool to solve free-boundary problems for fracture. Those free-boundary problems have been formulated in an idealized picture of fracture, which is viewed as the highly nonlinear stage of the Asaro-Tiller-Grinfeld (ATG) instability of a stressed solid surface \cite{Asaro:1972mz, Grinfeld:1986ly, PhysRevLett.71.1593}. In this picture, the crack dynamics is governed by a well-defined set of sharp-interface equations. In contrast to the other models, the crack tip scale is not set by the phase field interface thickness, but by a ``physical'' selection principle that can be rigorously derived from the analysis of the sharp-interface equations. This approach has close analogies with the dendritic growth problem, where the selected tip radius is of the order of the geometric means of the capillary and the diffusion length.
However, due to the different physics of the crack problem, the selection of the tip scale and growth velocity is different in details. For both dendrite and crack growth, the product of the tip radius and the growth velocity is fixed by the driving force (undercooling or applied elastic load, respectively), which determines the rate of energy dissipation. For dendritic growth, the steady state velocity scales proportionally to the ratio of heat or solute diffusivity and a microscopic capillary length, and is nontrivially selected through the anisotropy of surface tension. In contrast, for fracture, this velocity scales proportionally to the sound speed and is selected even for an isotropic fracture energy. Crack tip selection requires no adjustable parameters and naturally predicts propagation velocities appreciably below the Rayleigh speed. However, those predictions are generally limited by the strong assumptions made in formulating the sharp-interface equations governing the crack dynamics, and it is not yet clear how to relate those predictions to experiments for specific materials.

\begin{figure}
\begin{center}
\epsfig{file=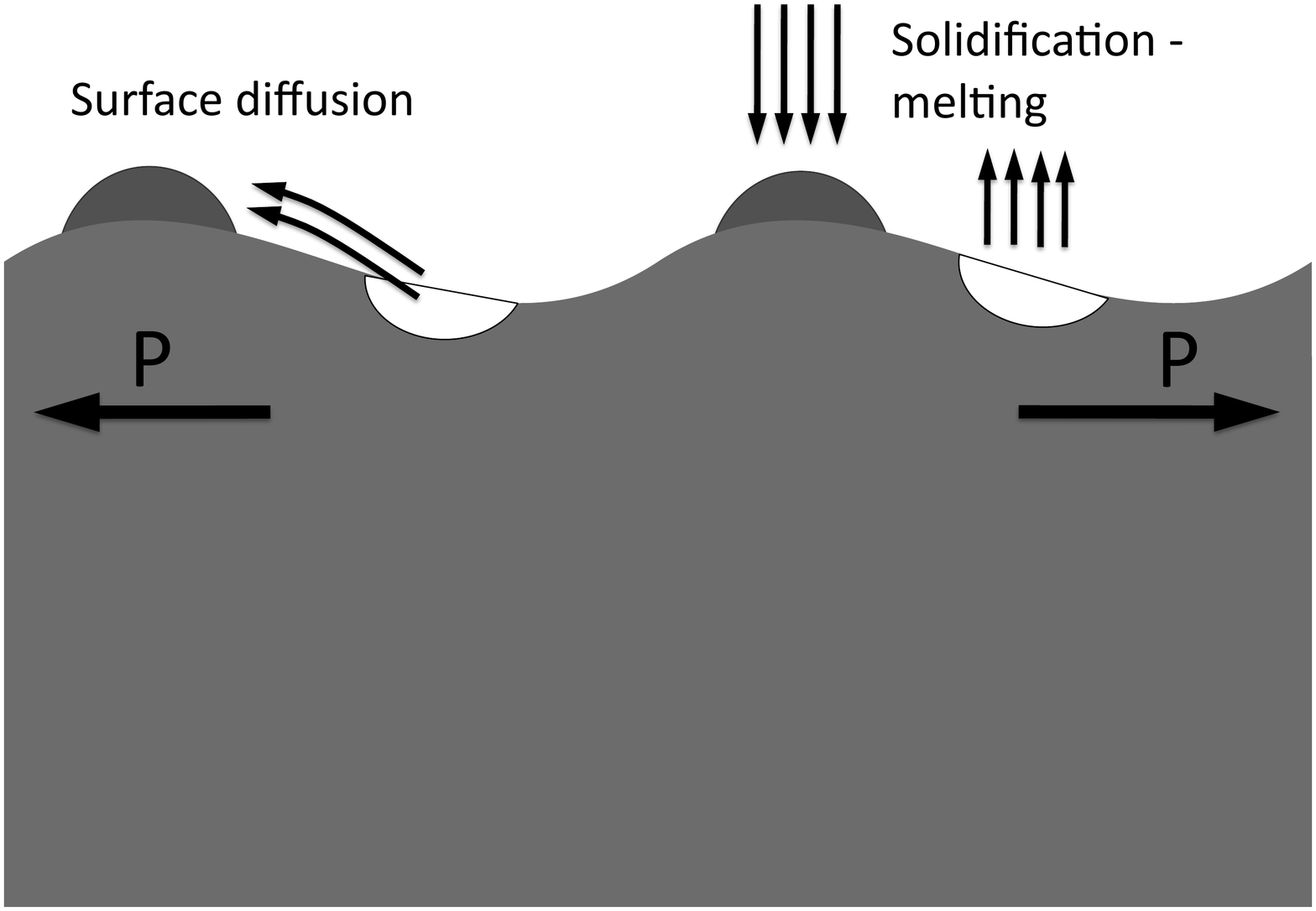, width=8cm}
\caption{The Asaro-Tiller-Grinfeld instability. The (two-dimensional) isotropic solid is stretched or compressed with a uniaxial stress $P$ along the interface, which leads to a morphological long-wave instability of the front. Material redistribution can be due to surface diffusion (left) or, if the solid is in contact with its melt, solidification and melting (right).}
\label{grinfeld}
\end{center}
\end{figure}

The starting point for this approach is based on the ATG instability, which predicts that a uniaxially strained solid surface is morphologically unstable \cite{Asaro:1972mz, Grinfeld:1986ly, PhysRevLett.71.1593}, see Fig.~\ref{grinfeld}.
The reason is that a deformation of the surface by a rearrangement of material can lead to an overall reduction of the total free energy.
Although the corrugation of the surface increases the surface free energy, the elastic energy can be reduced even more, since the stresses are released close to the interface.
The dynamics of this process is usually assumed to be either surface diffusion for a free surface or melting-solidification dynamics for a solid in contact with its melt at or close to the melting temperature.
Here we mention that an elastic deformation of the solid increases its free energy relative to the liquid, which can induce a stress-induced melting process.
For a perturbation of the interface contour $y(x,t)=A \cos kx \exp(\lambda t)$ the spectrum of the instability is for melting-solidification dynamics in linear approximation
\begin{equation} \label{grinfeld::eq1}
\lambda = K [k L_{ATG}-(k L_{ATG})^2], 
\end{equation}
with a kinetic coefficient $K$ for the interface kinetics and the Grinfeld length
\begin{equation}
L_{ATG}= \frac{E \gamma}{2(1-\nu^2) P^2},
\end{equation}
with the surface energy density $\gamma$, Young's modulus $E$ and the Poisson ratio $\nu$.
Notice that this lengthscale is -- apart from dimensionless factors -- the same as the Griffith length for crack growth, since both processes are the result of a competion between a release of elastic and an increase of surface (or fracture) energy.
As a result, long-wave perturbations are unstable, whereas short-wave corrugations are suppressed due to capillary effects, which are represented by the second term in Eq.~(\ref{grinfeld::eq1}).
The initial and late stage of the ATG instability has been modeled analytically, with sharp interface and phase field descriptions, see  e.g.~\cite{ChiuGao, 0295-5075-28-4-005, refId, PhysRevLett.82.1736, Kohlert:2003fk,PhysRevE.63.036117} and references therein;
\begin{figure}
\begin{center}
\epsfig{file=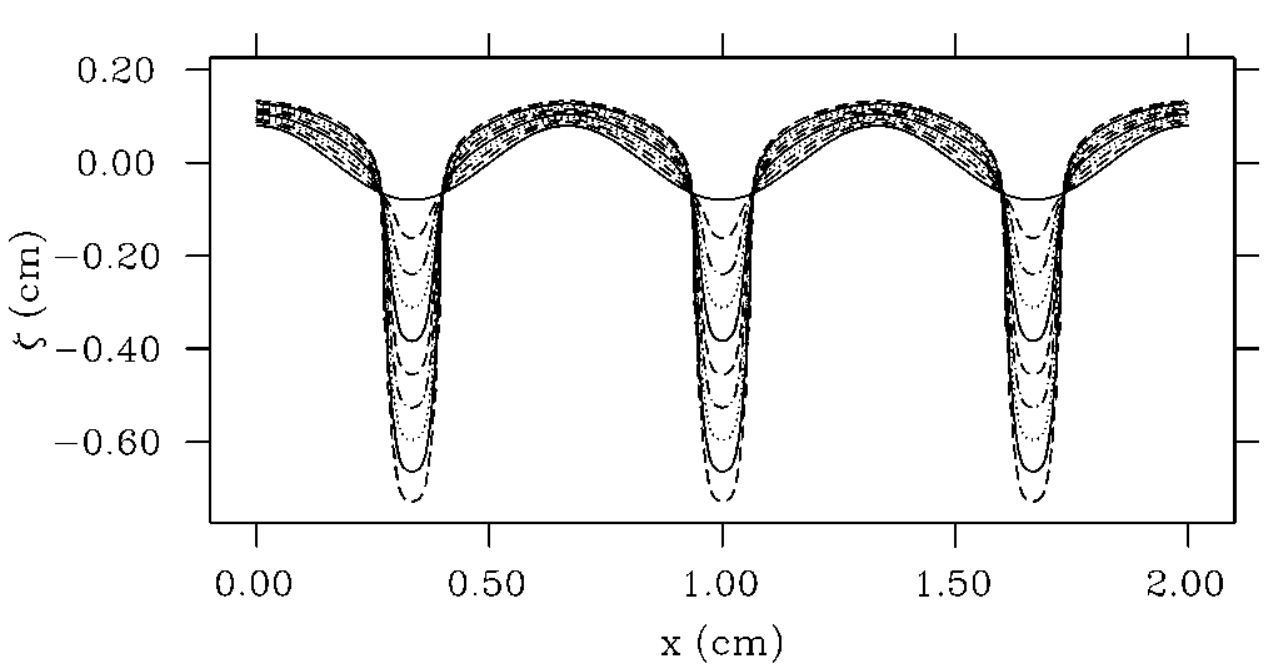, width=8cm}
\caption{Nonlinear evolution of the ATG instability, as obtained from phase field simulations. Taken from \cite{PhysRevE.63.036117}.}
\label{kassner}
\end{center}
\end{figure}
recently, also phase field models for surface diffusion reproduced the instability \cite{Raetz2006187, Gugenberger:2008fp}, and also phase field crystal investigations have been preformed \cite{wu:125408}.

In the late stage of the instability, the corrugations form deep grooves, which accelerate and evolve toward crack-like notches in the solid, see Fig.~\ref{kassner}.
In the framework of the static theory of elasticity, the tips of these ``cracks'' become arbitrarily sharp and reach an infinite velocity if no other cutoff mechanisms are provided (finite time cusp singularity).
This fast growth, apart from the missing regularization for static elasticity, therefore proposes a deeper link between the late stage of the ATG instability and crack growth. 

The first rigorous connection between cracks and the ATG instability was highlighted in \cite{PhysRevLett.81.5141}, where it was observed that the conditions for the occurrence of the instability, a uniaxial stress, are satisfied along the crack surfaces.
For a mode I crack with finite length, that it subjected to a tensile stress far away, the normal and shear stresses vanish on the crack surfaces, but the tangential stress is different from zero.
Therefore, the appearance of the instability is predicted for sufficiently long cracks, since the condition of fixed crack tips leads to ``quantization'' conditions for the spectrum of the instability, and at least one unstable mode has to fit on the straight crack.
The exact length threshold for the instability was later calculated numerically in \cite{PhysRevE.64.046120}, which shows that this long-wave instability can be expected for cracks that are already several times longer than the Griffith threshold, i.e.~for growing cracks.

Up to this point, the dynamics of the crack tip has been excluded from the considerations, and in \cite{PhysRevE.67.016112} a first model was proposed to describe the motion of the crack tip itself as a stress induced rearrangement of material.
There, in particular, the advancement of the crack was suggested to follow effectively a surface diffusion process.
The crack tip was introduced as a new degree of freedom in the sense that it has a spatially extended structure (finite tip radius), which removes the problem of the stress singularity.
Crack propagation therefore requires the redistribution of material if such a crack penetrates a solid.
If the excess material (the volume inside the crack) would have to be transported out of the crack completely, the process would be slow.
For a short diffusion path of the order of the tip scale, however, this is a conceivable mechanism for fast and steady state crack propagation.
This suggested growth mechanism, which is nothing else than the late stage of the ATG instability, therefore requires a scale selection of the tip, in order to prevent the aforementioned finite-time cusp singularity.
It was suggested that for fast crack growth the limitation to the Rayleigh speed provides the selection of the velocity, and together with the relationship between driving force and dissipation rate (the analogy of the Ivantsov relation), also the tip scale is selected.
Hence the only difference in comparison to the conventional treatment of the ATG instability is to take into account the finiteness of the sound speed, and this leads to a regularization of the theory.
In \cite{PhysRevE.67.016112}, the model was explored within the framework of a local stress model, which does cover potential tip instabilities.

\begin{figure}
\begin{center}
\epsfig{file=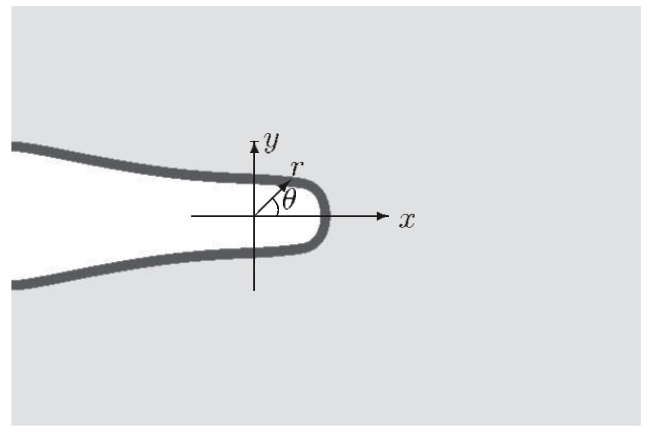, width=8cm}
\caption{Steady state growth of a crack with finite tip radius, determined through the interface kinetic coefficient and the sound speed. Taken from \cite{spatschek:015502}.}
\label{spatschek}
\end{center}
\end{figure}

A complete modeling of the problem using a phase field model was done in \cite{spatschek:015502}.
In contrast to \cite{PhysRevE.67.016112}, a phase-transition process is assumed at the crack surface, since it can be modeled more easily than the than the higher-order surface diffusion process \cite{Raetz2006187, Gugenberger:2008fp}.
The general selection principles, however, are analogous for both growth mechanisms.
There it is concluded that the suggested selection mechanism (finite sound speed) should indeed allow steady state growth and therefore prevent the finite time cusp singularity, thus steady state solutions with finite crack tip radius and velocity exist, see Fig.~\ref{spatschek}.
Also, crack branching is predicted for high driving forces due to a secondary ATG instability.
Since the crack blunts with higher driving force and the concentration of stresses in the tip region, unstable ATG modes should ``fit'' into the tip region beyond a critical driving force.
At that stage, the simulations were not yet quantitative, i.e.~the results for this model, which is based on sharp interface equations, still depended on the system size and the phase field interface thickness.

Due to the fact that the problem naturally contains several very different lengthscales, which are the macroscopic system size, the tip scale, the phase field interface thickness and the grid spacing for a numerical implementation, the required scale separation is difficult to achieve.
An alternative sharp interface description for the steady state regime was developed in \cite{pilipenko:015503}.
This approach has the advantage that it removes both the phase field interface thickness and the finite system size from the description.
The idea is that, in contrast to the singular stress field of a {\em sharp} crack, see Eq.~(\ref{eq1}), also higher order modes with a stronger divergency at the crack tip can be present for a {\em blunt} tip:
\begin{equation}
\sigma_{ij} = \frac{1}{\sqrt{2\pi r}} \left( K f_{ij}(\theta) + A_1 \frac{f_{ij, 1}(\theta)}{r} + A_2 \frac{f_{ij, 2}(\theta)}{r^2} + \ldots \right).
\end{equation}
Here, the angular dependencies $f_{ij, k}$ are again known functions, since they are eigenmodes of a straight cut, and the weight factors $A_k$ serve as expansion coefficients.
The higher order modes cannot be excluded, since the usual argument that they carry an infinite elastic energy is not valid for a crack with finite crack tip radius which serves as a cutoff.
Therefore, the stress field becomes a superposition of all these modes, and only the slowest decaying one is associated with the stress intensity factor, that is related to the driving force.
The other expansion coefficients $A_k$ are used to solve the linear elastic problem of traction free crack surfaces for a steady state crack.
Then the shape of the crack and its velocity are determined self-consistently for steady state solutions.
The results confirm the basic picture of the phase field simulations, i.e.~the existence of steady state solutions with a velocity appreciably below the sound speed and a finite tip radius.
Branching is indicated by a transition towards a crack tip with negative curvature in the steady state regime.
The results differ in detail from \cite{spatschek:015502}, and in particular the crack velocity is predicted as a partially decaying function of the driving force, which is a counterintuitive result.
Although the velocity dependence is rather weak, the applicability of the model to real materials remains an open issue.
Large-scale phase field simulations with a careful extrapolation to a full scale separation and infinitely large systems confirm these results \cite{spatschek:066111}.

In the preceding phase field and sharp interface models the crack surface is considered as the only source of dissipation, in agreement with the concept of small scale yielding.
It is however known that for many materials e.g.~plastic effects and the formation of defects play a crucial role in the vicinity of the crack tip.
As a first step towards an incorporation of such effects crack growth in viscoelastic media has been considered in \cite{Spatschek:2008qr}.
Here, the viscous damping coefficients introduce together with the surface diffusion coefficient a timescale, which finally provides selection for both the crack tip velocity and the tip radius.
In contrast to the inertia limited regime, the growth velocity is a monotonically increasing function of the driving force up to the point of crack branching.
Especially for mode I cracks, the contribution of the viscous damping turns out to be dominant in comparison to the dissipation at the crack front, and leads to a renormalization of the Griffith point;
below this apparent growth threshold (but above the literal Griffith point from energy balance), the crack grows only very slowly.
The higher the admixture of a mode III loading, the faster the crack propagates in the low driving force regime, which may indicate a front instability.
First attempts towards linking the inertia and bulk damping limited regimes is done in \cite{FracturePatternFleck}, employing both sharp interface and phase field methods.
Similarly, a crack tip scale selection can be induced by plastic yielding \cite{lo:027101}.

\section{Outlook}

Different phase-field approaches for fracture have been developed and analyzed to various degrees over the last decade. Crack propagation at low speed is governed essentially by macroscopic linear elasticity with traction free boundary conditions on the fracture surfaces. Therefore simple phenomenological phase field descriptions of fracture, such as the KKL model \cite{PhysRevLett.87.045501}, are generally able to describe this evolution quantitatively because the dynamics does not depend sensitively on details of the process zone physics. The analysis of such models has reproduced known crack propagation laws, such as the principle of local symmetry for isotropic media, and has provided a fertile ground for extending those laws to anisotropic media within the traditional energetic framework of continuum fracture mechanics. Furthermore, numerical examples to date have demonstrated the ability of those models to describe quantitatively crack kinking and oscillatory instabilities with biaxial loading and during thermal fracture.

For fast moving inertial cracks, phase field models have been able to reproduce crack branching instabilities seen experimentally. However, unlike for slow moving cracks, the threshold velocity for branching depends generally on details of the process zone description, which remains largely phenomenological. Therefore, while it is encouraging that this threshold is in the observed range of some fraction of the wave speed, it is not yet clear how to relate quantitatively phase field model predictions of dynamical branching instabilities to experimental observations. 
Phase field models that incorporate a more realistic description of the process zone are presently needed to make predictions more broadly applicable.
Inertial cracks have also been studied using a different class of phase field models that treat fracture as the late stage of the morphological instability of a stressed solid \cite{spatschek:015502}. Those models have the advantage of having a well-defined sharp-interface limit, which makes it possible to derive explicit scaling predictions for the crack velocity. Those predictions have highlighted interesting analogies and differences between the tip selection problems for cracks and dendritic crystals. However the physical description of fracture in those models is also too idealized to make quantitative predictions for real materials. 

Effects that are related to the discrete nature of the material, including lattice trapping effects, which produce a velocity gap and also affect macroscopically observable properties \cite{0295-5075-85-5-56002}, are not contained in the present phase field models.
Incorporating such effects in a continuum formulation presents a major challenge. Furthermore, in many materials crack propagation is accompanied by nonlinear elastic effects \cite{PhysRevLett.103.164301, PhysRevLett.101.264302, PhysRevLett.101.264301}, severe plastic deformation, as well as dislocation emission  that is believed to play a key role in the brittle-to-ductile transition. Phase field approaches that incorporate dislocations explicitly and the phase field crystal model have the potential to model those processes, but investigations of those models are still in very early stages for fracture. Another promising approach to tackle this transition is to modify the KKL model to incorporate a continuum description of plasticity. 

Finally, phase field models for crack growth have been studied so far only in two dimension, with the exception of Ref. \cite{PonsKarma2010}, where three-dimensional crack front instabilities under mixed mode loading are investigated. The computational cost for 3D simulations is very high, and eventually grid adaptivity and highly parallelized schemes have to be used to provide efficient algorithms.

\section*{Acknowledgements}

This work was supported by DOE grant DE-FG02-07ER46400 and the DOE sponsored Computational Materials
Science Network program.
R.S. acknowledges support of the German DFG grant SPP 1296 and the financial support from the industrial sponsors of ICAMS, ThyssenKrupp Steel AG, Salzgitter Mannesmann Forschung GmbH, Robert Bosch GmbH, Bayer Materials Science AG, Bayer Technology Services GmbH, Benteler AG and the state of North-Rhine-Westphalia.

\bibliographystyle{tPHM}
\bibliography{/Volumes/r.spatschek/Documents/Physics/BibDesk/references.bib}

\end{document}